\documentclass[twocolumn,aps,prb]{revtex4}
\usepackage{graphics,graphicx}
\usepackage{dcolumn}
\usepackage{bm}

\newcommand{\Ref}[1]{Ref.~\onlinecite{#1}}

\def\eb{\begin{equation}}   
\def\ee{\end{equation}}     
\def\ea#1{\begin{eqnarray} #1 \end{eqnarray}}   

\def\shro{Schr\"odinger}

\def\ra{\rightarrow}

\def\im{\text{Im}}

\def\of#1{\left(#1\right)}

\def\prtsq#1{{\partial^2 \over \partial {#1}^2}}

\def\eq#1{Eq.~(\ref{#1})}
\def\eqs#1#2{Eqs.~(\ref{#1}) and (\ref{#2})}

\def\sof#1{\left[ {#1} \right]}

\def\Dlt{\Delta}

\def\Pp{\Psi_+}
\def\Pm{\Psi_-}
\def\Ppm{\Psi_\pm}

\def\Ppmi#1{\Psi_{{#1}\pm}}
\def\Pmpi#1{\Psi_{{#1}\mp}}

\def\tshift{t_{\text{shift}}}

\def\Veff{V^{\text{eff}}}

\begin{document}

\title{Reconciling Semiclassical and Bohmian Mechanics: \\
IV. Multisurface Dynamics}

\author{Bill Poirier}
\affiliation{Department of Chemistry and Biochemistry, and
         Department of Physics, \\
          Texas Tech University, Box 41061,
         Lubbock, Texas 79409-1061}
\email{Bill.Poirier@ttu.edu}

\author{G\'erard Parlant}
\affiliation{Institut Charles Gerhardt, Universit\'e Montpellier 2, CNRS, \\ Equipe CTMM,
case courrier 1501, place Eug\`ene bataillon, 34095 Montpellier, France}
\email{Gerard.Parlant@univ-montp2.fr}

\begin{abstract}

In previous articles [J. Chem. Phys. {\bf 121} 4501 (2004),
J. Chem. Phys. {\bf 124} 034115 (2006), J. Chem. Phys. {\bf 124}
034116 (2006)] a bipolar counter-propagating wave decomposition,
$\Psi = \Psi_+ + \Psi_-$, was presented for stationary states
$\Psi$ of the one-dimensional \shro\ equation, such that the components
$\Ppm$ approach their semiclassical WKB analogs in the large action limit.
The corresponding bipolar quantum trajectories are classical-like and
well-behaved, even when $\Psi$ has many nodes, or is wildly oscillatory.
In this paper, the method is generalized for multisurface scattering
applications, and applied to several benchmark problems. A natural
connection is established between intersurface transitions and
$(+\!\leftrightarrow\!-)$ transitions.

\end{abstract}

\maketitle




\section{INTRODUCTION}
\label{intro}

This paper is the fourth in a
series\cite{poirier04bohmI,poirier06bohmII,poirier06bohmIII} investigating the
use of ``counter-propagating wave methods''
(CPWMs)\cite{poirier04bohmI,poirier06bohmII,poirier06bohmIII,poirier07bohmalg,babyuk04,wyatt}
for solving the time-independent \shro\ equation exactly. The basic idea is to
decompose the stationary wavefunction, $\Psi$, into a two-term,
or ``bipolar'' form,
\eb
     \Psi = \Pp +\Pm,  \label{psitot}
\ee
such that the $\Pp$ and $\Pm$ represent wave ``components'' moving in opposite
directions.  Although \eq{psitot} is exact, the particular bipolar decomposition
used is chosen to correspond to analogous approximate semiclassical wave
components.\cite{heading,froman,berry72}  For smooth
potentials, the latter are known to exhibit smooth and slowly-varying field
functions throughout the interaction region. The exact $\Ppm$ of \eq{psitot}
must behave similarly, at least in the classical limit, and in any case lead
to field functions that are much better behaved than for $\Psi$ itself. For
stationary scattering states, for instance, $\Psi$ necessarily exhibits oscillatory
interference in one or more asymptotic regions where the interaction potential $V$
is flat, whereas the asymptotic $\Psi_+$ and $\Psi_-$ behave as plane waves.
Thus, interference manifests not in the individual components, but arises
naturally from their linear superposition.

Apart from certain conceptual and theoretical advantages, the above picture
is particularly relevant for quantum trajectory methods
(QTMs),\cite{wyatt,lopreore99,mayor99,wyatt99,shalashilin00,wyatt01,wyatt01b,wyatt01c,wyatt02b,burghardt01b,bittner02b,hughes03}---i.e.,
trajectory-based numerical techniques for performing exact quantum dynamics
calculations, in a manner similar to classical simulations.\cite{frenkel}
QTMs that are based on standard Bohmian
mechanics\cite{madelung26,vanvleck28,bohm52a,bohm52b,takabayasi54,holland}
use a single-term or ``unipolar'' representation of the wavefunction, from which
the quantum trajectory evolution is determined. However, the Bohmian time evolution
equations are nonlinear, which can lead to radically different QTM behavior when
applied to the individual bipolar components of \eq{psitot}, rather than to $\Psi$
itself. In particular, the smooth field functions of the $\Ppm$ obviate the
``node problem''---i.e. the near-singularities in the quantum potential, $Q$,
leading to numerical instabilities in the quantum trajectory evolution when there is
substantial interference.\cite{babyuk04,wyatt,wyatt01b,trahan03,kendrick03,pauler04}
A more detailed discussion may be found in the previous articles cited above.

For one-dimensional (1D) stationary scattering state calculations, the bipolar
CPWM trajectories are actually {\em classical}, in a certain generalized
sense (Sec.~\ref{theory} and \Ref{poirier07bohmalg}). Quantum effects arise
not from a quantum potential, $Q$, impacting trajectory evolution, but
rather through dynamical coupling between the two components, $\Pp$ and $\Pm$.
This coupling, which is in essence proportional to the interaction potential,
induces $(+\!\leftrightarrow\!-)$ transitions---i.e., scattering from the
$\Pp$ incident/transmitted wave to the $\Pm$ reflected wave, and vice-versa.
In many respects, the situation is reminiscent of traditional multisurface
scattering theory, in which an off-diagonal diabatic coupling potential
induces a dynamical transition from one diabatic state to another. Indeed,
one somewhat compelling conclusion of the present
work---which deals with a bipolar CPWM treatment of multisurface dynamics---is
that in some sense {\em both types of transitions may be regarded as different
aspects of the same underlying phenomenon}. This ``unification'' may result in an
interesting cross-fertilization of ideas, e.g. in the area of classical
trajectory surface hopping (TSH) methodologies, originally designed
by Tully.\cite{tully71,tully90}

We are by no means the first to apply QTMs to the dynamics of electronic
nonadiabatic collisions; the first papers to do so were written several
years ago by Wyatt and coworkers.\cite{wyatt01,wyatt02b} It should be noted
that Wyatt's treatment is exact---at least formally---although approximate and/or
classical versions have also been developed.\cite{garashchuk05,garashchuk05b,burant00}
These methods are often compared to TSH, with which they (and the present approach)
have important differences. In TSH, a swarm of independent classical trajectories
evolve along a given potential energy surface; for each trajectory, a decision
is made whether to ``hop'' to a different surface, based upon a transition probability,
obtained by integrating coupled equations for the electronic amplitudes. In contrast,
the above multisurface methods do not involve any trajectory hopping
(the number of trajectories evolving on each electronic surface is, in fact,
 conserved) but transfer density and phase information from one state to
another in a continuous manner. Note that---like its single surface
counterpart---the multisurface QTM of Wyatt et al. is based on a unipolar
representation of the wavefunction, and is therefore also subject to numerical
instabilities due to interferences. Moreover, in the multisurface case,
interference arises not only from waves propagating in opposite directions
on the same surface, but also from waves transferring flux from one surface
to another.

It should also be stated that in the multisurface context, the idea
of applying a bipolar decomposition has been previously
considered. In particular, M. Alexander and coworkers adopted such a
scheme in the exact quantum solution of the close-coupling equations
using log-derivative propagation,\cite{alexander89,alexander91}
although their choice of \eq{psitot} decomposition does not avoid
oscillatory field functions, and is therefore not so useful for
QTMs. D. V. Shalashilin also used a bipolar decomposition to solve
the close-coupling equations (for inelastic scattering
applications), albeit only as a semiclassical
approximation.\cite{shalashilin93,shalashilin95}

The remainder of this paper is organized as follows. Sec.~\ref{theory} discusses
theoretical and algorithmic developments, both for single surface dynamics
calculations (Sec.~\ref{singlesurf}) and the multisurface case
(Secs.~\ref{multisurftheory} and~\ref{multisurfalg}). Note that the bipolar CPWM
algorithms used here require neither complex
scaling\cite{complexscaling78,reinhardt82,ryaboy94} nor absorbing
potentials\cite{seideman92a,riss93,poirier03capI,muga04}---a
decided advantage over other quantum scattering methods. Moreover, the scaling of
computational (CPU) effort is linear with the grid size, $N$, rather than
proportional to $N^3$. Results are presented in Sec.~\ref{results}.
For single surface applications
(Sec.~\ref{singleapp}), in addition to a test suite of wide-ranging applications
considered previously,\cite{poirier07bohmalg} we investigate the ability of the method
to compute scattering quantities to extremely high relative accuracy, or precision.
Several benchmark two-surface applications are considered in Sec.~\ref{multiapp},
including two of the Tully models.\cite{tully90} Concluding remarks are given
in Sec.~\ref{conclusion}.


\section{THEORY AND ALGORITHM DEVELOPMENT}
\label{theory}

\subsection{Single Surface Dynamics}
\label{singlesurf}

In several previous articles,\cite{poirier06bohmII,poirier06bohmIII,poirier07bohmalg}
an extremely accurate, efficient, and robust bipolar CPWM numerical algorithm was
developed for computing stationary scattering states of 1D single-surface systems. A
brief summary is presented here; further details can be found in the
above-referenced articles.

The algorithm is a time-dependent relaxation method, for which the
initial wavefunction is a left-incident/transmitted wave only, i.e.
$\Psi(x,t=0) = \Psi_+(x,t=0)$ and $\Psi_-(x,t=0) = 0$. Over time, a
reflected wave $\Psi_-(x,t)$ comes into being through interaction
region coupling due to the potential energy, and eventually
$\Psi(x,t) = \Psi_+(x,t)+\Psi_-(x,t)$ relaxes to the true stationary
eigenstate of desired energy, $E$, and left-incident boundary
conditions.  The solution components $\Ppm(x)$ behave as
(right/left) traveling plane waves in both asymptotes, except that
$\lim_{x\ra \infty} \Psi_-(x) = 0$. The asymptotic square
amplitudes, $\lim_{x\ra \infty}|\Psi_+(x)|^2$ and $\lim_{x\ra
-\infty}|\Psi_-(x)|^2$, are directly related to transmission and
reflection probabilities, respectively [\eq{Prefltrans}]. Through the
interaction region, the solution $\Ppm(x)$'s are exact quantum
analogues of a type of semiclassical WKB approximation resulting
from the ``generalized  Fr\"oman'' (F)
approach.\cite{poirier06bohmIII,poirier07bohmalg,froman} Amplitude and phase
functions for the $\Ppm$ components are expected to be smooth and
slowly varying, unlike those for the solution $\Psi$ itself.

The ``generalized'' aspect of the above methodology implies that we
are formally allowed to specify the ``classical'' trajectories as we
wish, using an effective potential $\Veff(x)$ of our own choosing,
rather than the actual potential $V(x)$. In practice, the method
requires that $\lim_{x\ra\pm\infty}\sof{V(x)-\Veff(x)}=0$, and works
best when $\Veff(x)$ is smooth and monotonic between the two
asymptotes. This ensures that all trajectories are devoid of turning
points, and that there is no asymptotic coupling between the two
$\Ppm$ components. Note that the $\Psi_+(x)$ trajectories move to
the right with velocity $v(x)= \sqrt{2\sof{E-\Veff(x)}/m}$, and the
$\Psi_-(x)$ trajectories move to the left with equal and opposite
velocity $-v(x)$. As per other QTM's, the trajectories give rise to
discrete moving grids, which carry local phase and amplitude
information for the appropriate $\Ppm$ component. There is no
quantum potential\cite{wyatt,holland} contributing to the
dynamics; all trajectories are (generalized) classical, and can be
determined {\em a priori}. Instead, quantum effects manifest via
$\Ppm$ coupling in the interaction region---as evidenced by the time
evolution equations (\Ref{poirier07bohmalg} Eq.~(16), or
\eq{Pdotgeneral} of this article with $f=1$). The coupling is
essentially proportional to the interaction potential, and can be
said to induce transitions from one $\Ppm$ component to the other.

In the previous papers, various numerical innovations were
introduced to improve performance and stability of the
trajectory-based time propagation of $\Ppm$, designed to deal
effectively with the unusual mixed boundary conditions [\eq{bcs}].
First, since the upper ($+$) and lower ($-$) grids are moving in
opposite directions, they cannot be commensurate at all times.
However, a scheme was introduced whereby these two grids are
commensurate at all times $t$ equal to a multiple of the ``shift
time,'' $\tshift$---the time it takes one grid point to travel to
the location vacated by its nearest neighbor. A constraint on
$\tshift$ is that it be a multiple of the propagation time step
size, $\Dlt$. Second, since grids are generally incommensurate,
interpolation must be used to evaluate the coupling contribution---a
procedure found to yield far better results when applied to a
{\em polar} (amplitude/phase) decomposition of the $\Ppm$'s,
rather than directly to the $\Ppm$'s themselves. A ``plane wave
propagator'' (PWP) approach was also employed, which effectively
treats the uncoupled part of the time evolution equations exactly,
and the coupled part using first-order Euler time integration.

Unfortunately, the PWP idea is found to be generally incompatible
with more efficient time integrators, such as fourth-order
Runge-Kutta (adaptive and non-adaptive), used here. Note that for
symmetric potential systems, it is possible to recast the time
evolution equations in such a way that the PWP contribution vanishes
(by introducing a time-evolving phase into the definition of the
wavefunction). In order to ascertain the extent to which PWP
actually improves performance, some preliminary numerical tests were
conducted for the Eckart A system (Sec.~\ref{results}) with $E=V_0$.
In particular, an efficiency comparison was made between the
original PWP algorithm and the ``phase-modified'' version---both
using first-order Euler propagation for the non-PWP contribution.
For an identical set of numerical parameters, both methods performed
about equally well, vis-\`a-vis the accuracy of the computed
reflection and transmission probabilities---although the
phase-modified version executed roughly $30\%$ faster, owing to less
required computation.  As a second, more realistic test, the
phase-modified algorithm with fourth-order Runge-Kutta was compared
to a straight fourth-order Runge-Kutta treatment of all terms in the
original evolution equations (including PWP terms). Both methods
again performed roughly equally well with respect to computed
accuracy, with slight performance differences depending on the
desired level of accuracy. Moreover, both methods are far superior
to the first-order Euler methods, enabling $\Dlt$ values that are
orders of magnitude larger, without compromising numerical accuracy.
Only the straight Runge-Kutta method can be generalized for
asymmetric potential systems, however, and is introduced here as the
general method of choice.

We also introduce an improved method for dealing with extremal trajectories
that stray outside the range of interpolation: instead of ignoring the
coupling contribution altogether,\cite{poirier07bohmalg} we now simply use
values for the polar field quantities obtained from the nearest extremal grid
point---although it should be emphasized that the true coupling contribution
is vanishingly small in the asymptotic limit. Also, even within the range of
interpolation, the interpolated density value may become slightly negative
when the density absolute values are very small; in that case, we now reset
the interpolated value to zero.

\subsection{Multisurface Dynamics: Theory}
\label{multisurftheory}

We now generalize the theory for the 1D multisurface case. Note that only
a brief summary is provided here, as the full derivation follows closely
that of Refs.~3 and~4, which should be consulted for further details.

Let $f$ be the number of electronic states considered, with no restrictions
on intersurface coupling. A diabatic-like time-independent matrix
\shro\ equation is presumed, of the form \eb \tilde{H} \cdot \vec
\Psi = E \vec \Psi, \label{multiSE} \ee where
$\{\Psi_1,\Psi_2,...\Psi_f\}$ comprise the vector components
(associated with each of the $f$ diabatic states) of the nuclear
wavefunction, $\vec\Psi$, and \eb \sof{\tilde{H}}_{i,j} =
-\delta_{i,j} \of{{\hbar^2\over 2m}} \prtsq x + V_{i,j}(x)
\label{multiH} \ee are the components of the $f\times f$ Hamiltonian
operator matrix, $\tilde{H}$, with $i\le f$ and $j\le f$ labeling
diabatic states.

The $V_{i,j}(x)=V_{j,i}(x)$ are the diabatic potential energy
curves, with the $i \ne j$ case denoting the coupling potentials. In
order to ensure that coupling vanishes in the asympotic limits (required
to obtain asymptotic scattering waves with correct boundary
conditions),\cite{poirier06bohmIII,poirier07bohmalg}
we must have $\lim_{x\ra \pm \infty} V_{i\ne j}(x) = 0$. However, the
asymptotic values for the diagonal potentials, $V_{i,i}(x)$, are
allowed to be completely arbitrary, and in particular, need not be
symmetric. Left and right asympotic values are denoted $V_{iL} =
\lim_{x\ra - \infty} V_{i,i}(x)$ and $V_{iR} = \lim_{x\ra \infty}
V_{i,i}(x)$, respectively. For purposes of generating trajectories
for motion on the $i$'th diabatic state, a suitable family of $f$
effective potentials, $\Veff_i(x)$, are introduced, such that
$\lim_{x\ra \pm \infty} \sof{V_{i,i}(x) - \Veff_i(x)} = 0$ (to
ensure against asymptotic coupling), and $\Veff_i(x) \le
\max(V_{iR}, V_{iL})$ for all $x$ (to avoid turning points, as
discussed in Sec.~\ref{singlesurf}).

Assuming that each component of $\vec \Psi$ is decomposed into its
own bipolar expansion, \eb \Psi_i(x) = \Psi_{i+}(x) + \Psi_{i-}(x),
\label{multibipolar} \ee we clearly need $2f$ independent
differential equations [and associated boundary conditions
(Sec.~\ref{multisurfalg})], to uniquely specify the solution
decomposition of \eq{multibipolar}. One half of these are already
provided in \eqs{multiSE}{multiH}. For the remaining $f$ equations,
it is natural to apply the generalized $F$ approach
(Sec.~\ref{singlesurf}) in component-wise fashion. In particular,
for a single-surface system, the generalized F approach\cite{froman}
provides the following as the second (after the usual \shro\ equation)
independent differential condition,
\eb
     \Psi' = -{v' \over 2v} \Psi +
          {i m v \over  \hbar} \of{\Psi_+ - \Psi_-} \label{FPprimeB}
\ee
(where the prime denotes spatial differentiation), as discussed in more
detail in \Ref{poirier07bohmalg} [Eq.~(11)]. For the multisurface
generalization, we simply apply \eq{FPprimeB} to each diabatic state
separately, to obtain
\eb
     {\Psi_i}' = -{{v_i}' \over 2v_i} \Psi_i +
          {i m v_i \over  \hbar} \of{\Psi_{i+} - \Psi_{i-}}, \label{Fprime}
\ee
where $v_i(x) = \sqrt{2\sof{E-\Veff_i(x)}/m}$ are the trajectory velocities
for the $i$'th diabatic state (as determined by the effective potential $\Veff_i$).
Note that \eq{Fprime} does not depend at all on the $V_{i\ne j}$ as appropriate,
i.e. all intersurface coupling should arise through the \shro\ \eq{multiSE}, itself,
if $(+\!\leftrightarrow\!-)$ transitions for single- and multi-surface applications
are to be treated equally.

By combining \eq{multiSE} with \eq{Fprime}, we can derive
expressions for the first spatial derivatives of each of the $2f$
bipolar CPWM components, i.e. the ${\Psi_{i\pm}}'$, directly in
terms of the undifferentiated component quantities. This yields
somewhat complicated results, analogous to \Ref{poirier07bohmalg} Eq.~(12),
which are excluded here for the sake of brevity. By following the
procedure described in detail in \Ref{poirier06bohmIII} and \Ref{poirier07bohmalg}
[basically, constructing a convective term to get rid of first-order spatial
derivatives of the wavefunction in the hydrodynamic frame, and introducing
an explicit time dependence via $\partial \Ppm /\partial t = -(i/\hbar) E \Ppm$],
we are then led to the following coupled hydrodynamic (Lagrangian)
time evolution equations:
\ea {
    \lefteqn{{d \Ppmi i \over d t}  =} \hspace{3.0in}\nonumber\\
     \sof{ \pm {1\over 4} v_i \of{{{\Veff_i}' \over E-\Veff_i}}
                  + {i \over \hbar} \of{E - V_{i,i} - \Veff_i + C_i} } \Ppmi i & &\nonumber \\
      - {i \over \hbar} \sof{V_{i,i} - \Veff_i - C_i} \Pmpi i
                  - {i \over \hbar} \sum_{j\ne i}^f V_{i,j} \of{\Psi_{j+} + \Psi_{j-}},& &\label{Pdotgeneral}}
where,
\eb
C_i = \of{\hbar^2 \over 2m} \sof{{5 \over 16} \of{{{\Veff_i}'\over E-\Veff_i}}^2 +
                                        {1 \over 4} \of{{{\Veff_i}''\over E-\Veff_i}} }
                  \label{Cex}
\ee
For pedagogical purposes, we also consider the asympotically
symmetric special case with $V_{iL}=V_{iR}=0$ and $\Veff_i(x)=0$ for
all $i$:
\eb
     {d \Ppmi i \over d t} =  {i \over \hbar} \sof{E\Ppmi i
                  - \sum_{j=1}^f V_{i,j} \of{\Psi_{j+} + \Psi_{j-}} }
                           \label{Pdotspecial}
\ee
Note: in \eq{Pdotspecial}, $j$ ranges over all values, including
$j=i$.

The time evolution equations of the preceding paragraph---best
exemplified by \eq{Pdotspecial}---offer a coherent, unified picture
of scattering theory that places both intersurface transitions and
$(+\!\leftrightarrow\!-)$ transitions on a near-equal footing. Note
first that coupling of all types vanishes in both asymptotic limits,
resulting in uncoupled, scattering plane wave dynamics in these
limits for all $\Ppmi i$, as desired. In particular, $\Ppmi i
(x\!\ra\!\pm\infty,t)$ is a plane wave propagating to the
(right/left) with uniform speed $v_{iL/R} =
\sqrt{2\sof{E-V_{iL/R}}/m}$. It is the various potential
interactions, operating in the interaction region, that induce
transitions of all kinds among the $2f$ bipolar CPWM components,
$\Psi_{i\pm}$. The off-diagonal potentials $V_{i\ne j}$ are
responsible for transitions between diabatic states $i$ and $j$,
whereas the diagonal potentials $V_{i,i}$ induce transitions from
reactive (transmitted) $\Psi_{i+}$ to non-reactive (reflected)
$\Psi_{i-}$ components (and vice-versa). Apart from this
distinction, both types of transitions manifest similarly in the
equations of motion.

As in the previous work,\cite{poirier06bohmIII,poirier07bohmalg} the time evolution
equations above give rise to some elegant flux properties, which
again treat both types of transitions (or probability flow) on a
near-equal footing. Let $j_{i\pm} = \pm v_i \rho_{i \pm}$ be the
flux for component $\Ppmi i$, defined in terms of the generalized
trajectory velocities $v_i$, and the component probability
densities, $\rho_{i\pm} = |\Ppmi i|^2$. We also find it convenient
to introduce new composite labels, $\alpha = (i,\pm)$ and $\beta =
(j,\pm)$, to label individual $\Ppmi i$ components. Note that the
$\pm$ values are {\em independent} for $\alpha$ and $\beta$, so that
each of these two component labels can take on $2f$ distinct values.
Using \eq{Pdotgeneral}, and transforming to Eulerian (partial) time
derivatives, we obtain the following flux relation:
\eb
{\partial \rho_\alpha \over \partial t}  =  - {j_\alpha}' +
          \sum_{\beta \ne \alpha} {\partial \rho_{\text{cpl}}^{\alpha \leftarrow \beta}
          \over \partial t}, \label{fluxrel}
\ee
where
\eb
{\partial \rho_{\text{cpl}}^{\alpha \leftarrow \beta} \over \partial t}
           =  {2 \over \hbar}\sof{V_{i,j}-\delta_{i,j}\of{\Veff_i + C_i}}
          \im \sof{{\Psi_\alpha}^* \,\Psi_\beta}, \label{rhodotcpl}
\ee
and $\partial \rho_{\text{cpl}}^{\alpha \leftarrow \beta}/ \partial
t$ represents the rate of probability density flow from component
$\beta$ to $\alpha$.

Equations~(\ref{fluxrel}) and~(\ref{rhodotcpl}) above emphasize the essential similarities
between the two types of transitions---particularly for the $V_{iL} =V_{iR}=0$ case, for
which the $\delta_{i,j}$ contribution in \eq{rhodotcpl} vanishes.
More importantly, however, we find that
$\partial \rho_{\text{cpl}}^{\alpha \leftarrow \beta}/ \partial t = -
\partial \rho_{\text{cpl}}^{\beta \leftarrow \alpha}/ \partial t$, which implies
the nontrivial combined continuity relation,\cite{poirier06bohmIII,poirier07bohmalg}
\eb
     {\partial \sum_\alpha \rho_\alpha \over \partial t} = - \sum_\alpha {j_\alpha}'.
     \label{contcond}
\ee
Equation~(\ref{contcond}) further implies that the total probability integrated
over all $2f$ components is conserved over time---a desirable but nontrivial result,
given that $\Psi_{i+}$ and $\Psi_{i-}$ are not true ``components'' (orthogonal
complements) in the way that $\Psi_i$ and $\Psi_{j\ne i}$ are.

The above discussion regarding time evolution properties pertains to
all times, $t$---not just the $t\ra \infty$ limit, where $\vec\Psi$
approaches the exact stationary solution. In the latter limit,
however, it is clear that \eq{contcond} must be zero, so that the
stationary solution must satisfy ${j_+}' = - {j_-}'$, where $j_\pm =
\sum_{i=1}^f j_{i\pm}$ represents the total flux moving in the
(right/left) directions. For the $V_{iL} =V_{iR}=0$ case in
particular, this implies that ${\rho_+}' = {\rho_-}'$, where
$\rho_\pm = \sum_{i=1}^f \rho_{i\pm}$. In other words, the sum of
all $+$ densities is equal to the sum of all $-$ densities, apart
from a constant---a nice generalization of \Ref{poirier06bohmIII} Eq.~(15).

\subsection{Multisurface Dynamics: Numerical Algorithm}
\label{multisurfalg}

To implement the above time evolution equations numerically, we must
first discuss boundary conditions and initial value conditions.
Without loss of generality, we may assume an incoming wave that is
left-incident on the diabatic state $i=1$. Probability flow from
$\Psi_{1+}$ to the other CPWM components occurs only in the
interaction region, and then propagates outward towards the
appropriate asymptotes. Adopting the usual normalization convention
thus leads to the following boundary conditions: \ea{
          \Psi_{1+}(x\ra -\infty)& = &
              \exp\sof{+i \of{{\sqrt{2 m (E-V_{1L})}\, x \over \hbar}-\phi}}
               \nonumber \\
          \Psi_{(i>1)+} (x\ra -\infty)& = & 0 \label{bcs} \\
          \Psi_{i-} (x\ra \infty)& = & 0 \nonumber }
In \eq{bcs} above, $\phi$ is a time-dependent phase factor.

As for the initial value, we take $\Psi_\alpha(x,t=0) = 0$ for all
$\alpha$ except $\alpha = (1,+)$, for which the basic WKB solution
is used, as described in \Ref{poirier07bohmalg} Eq.~(18). In the $t\ra
\infty$ limit, the resultant numerical solution $\Psi_{i\pm}$'s can
be directly consulted to obtain transmission probabilities
[associated with  $\rho_{i+}(x\ra\infty)$] and reflection
probabilities [associated with  $\rho_{i-}(x\ra-\infty)$], as per
\eq{Prefltrans}.

In most respects, the numerical algorithm employed for the time evolution
is identical to that discussed in Sec.~\ref{singlesurf}. One complicating factor
is that the trajectories [determined from $v_i(x)$] are obviously completely
different from one diabatic state, $i$, to the next. In a single-surface
calculation, the $+$ and $-$ grid points move in opposite
directions. Thus, the two grids must be incommensurate for most times,
although a simple scheme is used to ensure commensurate grids when $t$
is a multiple of the shift time, $\tshift$. Starting at the left edge
of the grid, $x_L$, one propagates a {\em single} trajectory, $x(t)$.
The initial grid for both $+$ and $-$ is then taken to be the values
$x_k = x(t = k \tshift)$ for nonnegative integers, $k$ such that
$x_k < x_R$, where $x_R$ is the right edge of the grid.\cite{poirier07bohmalg}

In the multisurface case, the trajectories at a given point $x$ move
with different speeds for different $i$ values, thus rendering it
impossible for grids to be commensurate across diabatic states,
even at regular time intervals. A reasonable procedure, however, is
to apply the above in component-wise fashion, i.e. by generating a
family of $f$ trajectories, $x_i(t)$. If the same $\tshift$ value is
used for all $i$, then at intervals $t= k \tshift$, the $i+$ and
$i-$ grids will be commensurate with each other, for all $i$ (but
still not {\em across} $i$ values). The nonuniform grids generated
in this fashion will be denser where $v_i$ is smaller---which is
physically reasonable, and in any event is also true
in the single-surface case. To evaluate the intersurface coupling
contribution in \eq{Pdotgeneral}, a polar intersurface grid interpolation
scheme is used, exactly analogous to that used for
$(+\!\leftrightarrow\!-)$ coupling in the single-surface case.
Note that all grids extend over the full coordinate range, i.e.
roughly from $x_L$ to $x_R$.


\section{RESULTS}
\label{results}

In this section, we discuss the application of the numerical
algorithms described in Sec.~\ref{theory} to a variety of model
systems. The mass $m=2000$ a.u. was used in all cases. Also,
fourth-order Runge-Kutta time integrators were employed, as was
natural spline interpolation, unless stated otherwise. For the
symmetric, single-surface potentials, the phase-modified algorithm
was used. A thorough and detailed convergence study was performed
for each system, by varying each of the convergence parameters in
turn, and monitoring associated changes to the computed reflection
and transmission probabilities, $P_i^{\text{refl}}$ and
$P_i^{\text{trans}}$. Unless stated otherwise, the precise
convergence procedure followed is that described in \Ref{poirier07bohmalg},
with one exception: since ``oscillatory error'' was always found to
be negligible, reflection and transmission probabilities are defined
here to be \ea{
     P_i^{\text{refl}} & = &\of{v_{iL}/v_{1L}}|\Psi_{i-}(x_L)|^2 \nonumber \\
     P_i^{\text{trans}}& = &\of{v_{iR}/v_{1L}}|\Psi_{i+}(x_R)|^2, \label{Prefltrans} \\
     \text{where}\quad v_{iL/R} & = & \lim_{x\ra\mp\infty} v_i(x),\nonumber }
rather than as described in \Ref{poirier07bohmalg}.

\subsection{Single Surface Applications}
\label{singleapp}

As fourth-order Runge-Kutta integrators have not been previously
combined with the numerical innovations of \Ref{poirier07bohmalg}
(Sec.~\ref{singlesurf}), our first task is to evaluate the numerical
efficacy of the present approach for single surface systems. In
particular, we apply the method to a ``test suite'' of 1D benchmark
applications\cite{poirier06bohmIII,poirier07bohmalg} exhibiting a range of different
attributes with regard to tunneling, barrier height and width,
exoergicity, existence of reaction intermediates, and desired level
of computational accuracy. In doing so, the robustness and stability
of the method will be evaluated, as well as the numerical
efficiency, as compared with the previous codes.

Only a brief description of the individual test suite applications will be
provided here; for additional details, the previous papers may be consulted.
The five test suite potentials are as follows:
\begin{itemize}
\item{{\bf Eckart A:} short, narrow Eckart barrier (height $V_0 \approx 0.0018$ hartree)}
\item{{\bf Eckart B:} tall, wide Eckart barrier (height $V_0 = 0.011$ hartree)}
\item{{\bf Uphill ramp:} tanh potential, barrierless and monotonic
                        ($V_R-V_L \approx 0.0018$ hartree)}
\item{{\bf Barrier ramp:} Eckart $+$ tanh potential, asymmetric with barrier
                          ($V_R-V_L \approx 0.0018$ hartree)}
\item{{\bf Double barrier:} symmetric double barrier with reaction intermediate well
                           (barrier height $V_0 \approx 0.0018$ hartree)}
\end{itemize}
Results are presented in Table~\ref{pottab}.

\begin{table*}
\caption{\label{pottab}Parameters used for bipolar CPWM scattering calculations for
single-surface  test suite applications. All units are atomic units,
except Row 12, the required CPU time in seconds on a 2.60 GHz Pentium CPU.
Last digit of computed reflection and transmission probabilities
(Rows~8 and~9) is uncertain.}
\begin{ruledtabular}
\begin{tabular}{ccccccccc}
Quantity & \multicolumn{8}{c}{Single-surface test suite applications} \\ \cline{2-9}
   and      & \multicolumn{2}{c}{Eckart A, $E=\,V_0$} & \multicolumn{3}{c}{Eckart B, $E=$} &
               Uphill & Barrier & Double \\ \cline{2-3} \cline{4-6}
 symbol     & low acc. & high acc. & $V_0$ & $0.4\,V_0$ & $0.1\,V_0$ &
              ramp & ramp & barrier \\ \hline
energy,          $E$                & 0.001823 & 0.001823  & 0.011   & 0.0044    & 0.0011     & 0.0023  & .0023   & 0.0014 \\
grid size,       $N$                & 20       &  800      & 25      & 61        & 110        & 19      & 15      & 20     \\
left edge,       $x_L$              & -2.0     & -4.0      & -2.6    & -3.5      & -3.5       & -1.5    & -1.5    & -2.2   \\
right edge,      $x_R$              &  2.0     &  4.0      & 2.1     & 4.0       &  3.5       & 2.2     & 2.0     & 2.2    \\
time step,       $\Dlt$             & 156      &  2.47     & 19.7    & 14.9      & 30.62      & 107     & 125     & 196    \\
\# steps,        $\tshift/\Dlt$     & 1        &  3        &  3      &  4        & 2          & 2       & 2       & 1      \\
max time,        $t_{\text{max}}$   & 3899     & 11867     & 2893    & 43978     & 428621     & 5792    & 7000    & 39143  \\
\hline
computed         $P^{\text{refl}}$  &.28348   &.283...3842 & .45967  &           &            & .02391  & .45455  & .7958  \\
computed         $P^{\text{trans}}$ &.71665   &.716...6154 & .537    & 1.560(-5) & 9.923(-10) & .97595  & .54562  & .2053  \\
exact            $P^{\text{refl}}$  & \multicolumn{2}{c}
                                          {0.283358063869} & .459605 &           &            & .023901 &         &        \\
exact            $P^{\text{trans}}$ & \multicolumn{2}{c}
                                          {0.716641936131} & .540395 & 1.559(-5) & 9.920(-10) & .976099 &         &        \\
\hline
CPU time (s)                        & 0.0045   & 36        & 0.035   & 1.57      & 15         & 0.010   & 0.008   & 0.055  \\
\end{tabular}
\end{ruledtabular}
\end{table*}

In the first investigation, we computed $P^{\text{refl}}$ and
$P^{\text{trans}}$ for each of the five test suite potentials above,
at an energy $E \approx V_0$ or $E\approx (V_R-V_L)$
(Table~\ref{pottab}, Row 1) chosen to yield both substantial
reflection and transmission. The results are presented in Columns
II, IV, and VII--IX of Table~\ref{pottab}. The computed
$P^{\text{refl}}$ and $P^{\text{trans}}$ values (Table~\ref{pottab},
Rows~8 and~9) were both converged to an absolute accuracy of around
$10^{-4}$ (last digit uncertain). For the Eckart~A system (Column
II), all parameters were taken directly from \Ref{poirier07bohmalg},
except for the fixed (non-adaptive) Runge-Kutta time step size,
$\Dlt$, which was reconverged. For all other applications, a full
reconvergence of all parameters was performed, as discussed above.
For some systems, a direct comparison can be made between computed
and exact (Rows~10 and~11) $P^{\text{refl}}$ and $P^{\text{trans}}$
values. For the Eckart systems, the exact results are known
analytically.\cite{ahmed93} For the uphill ramp system, ``exact''
values were obtained from a much more accurate numerical
calculation, converged to at least eight significant digits.

In comparison with the corresponding calculations in
\Ref{poirier07bohmalg}, the converged parameter values are for the most
part very similar, as expected. The glaring exception, of course, is
the time step size, $\Dlt$, whose values are found here to be {\em
much} larger than in \Ref{poirier07bohmalg}, owing to the use of
fourth-order time integrators. More significant, however, is the
fact that the present $\Dlt$ values are substantially larger than
those of \Ref{poirier06bohmIII}, which {\em also} employed non-adaptive
fourth-order Runge-Kutta integration. For Eckart A for instance
(Column II), the respective $\Dlt$ values (in atomic units) are as
follows: \Ref{poirier07bohmalg}, 0.156; \Ref{poirier06bohmIII}, 10.0; present work,
156. The latter value is indeed very large for the level of accuracy
achieved, and can be attributed to the numerical improvements
introduced in \Ref{poirier07bohmalg}. Remarkably, an even larger $\Dlt$
could in principle have been used here, but the value given is
already so large that $\Dlt = \tshift$. Thus, over a single time
step, each grid point moves to a neighboring site---resulting in an
effectively ``fixed'' grid that obviates both spatial
differentiation\cite{poirier06bohmIII,poirier07bohmalg} {\em and} interpolation.
With regard to CPU effort, the Eckart A calculation performed here
required 4.5 ms on a 2.60 GHz Pentium CPU, as compared with 1.7 s
for \Ref{poirier07bohmalg}. This reduction in CPU time is not as large as
the $1000 \times$ factor predicted by the increase in $\Dlt$, owing
to the fact that each Runge-Kutta iteration requires additional
computation, but is nevertheless very substantial; the present
calculation is also orders of magnitude faster than the
corresponding \Ref{poirier06bohmIII} calculation.

In addition to the above investigation, we also examined the present
method's ability to provide extremely accurate results. This was
done in two distinct ways. First, a very deep tunneling study of the
Eckart~B system was performed, as presented in Table~\ref{pottab},
Columns V and VI. In effect, such a study evaluates absolute
accuracy rather than precision---e.g. Column VI, for which the
transmission probability is around $10^{-9}$, but is computed to
only three significant digits. The method clearly works well in this
context---although the greater the absolute accuracy level, the less
the improvement relative to \Ref{poirier07bohmalg}. For the $E = 0.1\,V_0$
case, for instance, $\Dlt$ increases by only a factor of $25
\times$, which is less than $\tshift$, but only by a factor of two
(Table~\ref{pottab}, Row 6). Second, In order to evaluate {\em
precision} rather than absolute accuracy, we also performed an
extremely accurate calculation of the $E=V_0$ Eckart~A system, for
which both $P^{\text{refl}}$ and $P^{\text{trans}}$ were computed to
{\em eleven} significant digits (Table~\ref{pottab}, Column III). As
compared with Column II, the time step is reduced to an extent that
is quantitatively consistent with a fourth-order integration method.
The corresponding increase in grid size is difficult to predict
quantitatively, but interestingly, is found to be such that the
$(\tshift/\Dlt)$ ratio is again not far from one. Even for this
extremely challenging application, the total CPU time required is
only 36 seconds.

Despite the substantial breadth of the test suite applications considered,
in all cases, the $(\tshift/\Dlt)$ ratios are on the order of one
(Table~\ref{pottab}, Row 6)---a situation that bears further scrutiny.
Accordingly, we conducted an additional calculation of the Eckart A
system using {\em adaptive} (Cash-Karp) fifth-order Runge-Kutta
integration\cite{press}---the idea being that an adaptive method should automatically
find the most appropriate time step sizes to use. In fact, using parameters
similar to Column II, and specifying a target accuracy of $5\times 10^{-5}$,
we found the resultant $\Dlt$ values to be almost always equal to $\tshift$, except
for the first few time steps, which had much smaller $\Dlt$ values.  The resultant computed
transmission and reflection probability errors (relative to exact analytical values)
conform to the target accuracy level. One might conclude from the above
that the corresponding {\em non}-adaptive calculation using fixed $\Dlt = \tshift$
would yield similar results. In fact, the latter calculation
results in a near identical $P^{\text{trans}}$ value, but a $P^{\text{refl}}$
with an error of $1.0 \times 10^{-3}$. Thus, the initially small time steps appear
to play a vital role, and also argue cogently in favor of adaptive time
integration schemes.


\subsection{Multisurface Applications}
\label{multiapp}

We also performed numerical calculations for a variety of benchmark multisurface
applications for $f=2$, as described below.
\begin{itemize}
\item{{\bf Pure coupling:} asymptotically symmetric, with $V_{ii}(x)=0$ and Gaussian coupling.}
\item{{\bf Tully model 1:} crossed ramp potentials, $V_{ii}(x)$, with Gaussian coupling.}
\item{{\bf Tully model 2:} double avoided crossing, with $V_{22}(x)$ a potential well.}
\end{itemize}
These systems exhibit a very broad range of attributes, especially
with respect to energy scale.

\subsubsection{Pure coupling system}
\label{pcsys}

The two-surface pure coupling system is defined via
\ea{
     V_{11}(x) & = & V_{22}(x) = 0 \nonumber \\
     V_{12}(x) & = & V^{12}_0 \exp[-\alpha (x-x_0)^2], \label{pcpot}
}
with the parameter choices $V^{12}_0 = 150\,\text{cm}^{-1}$,
$x_0=0$, and $\alpha = 1$ a.u. This system represents an extreme
case, in that there are no interaction potentials $V_{i,i}$ to
induce direct $(+\!\leftrightarrow\!-)$ transitions. Instead, any
reflection must arise indirectly, from a successive pair of
intersurface transitions. As is clear from \eq{pcpot}, the pure
coupling system above is asympotically symmetric, implying that we
can use uniform trajectories (which are also classical trajectories)
and \eq{Pdotspecial}.

The energy $E=  100\,\text{cm}^{-1} = (2/3)V^{12}_0$ is chosen to
result in substantial probabilities for both scattered and
unscattered wave components.  Numerical calculations were performed
using the algorithm of Sec.~\ref{multisurfalg}, with adaptive
Cash-Karp integration as discussed in Sec.~\ref{singleapp}. All
reflection and transmission probabilities were converged to an
accuracy of $10^{-5}$ or better, using the following parameter values:
$N=61$; $x_{L/R} = \mp 3$ a.u.; $t_{\text{max}} = 50 000$ a.u.; Runge Kutta
tolerance level $\epsilon = 10^{-6}$.

\begin{figure}
\includegraphics[scale=0.85]{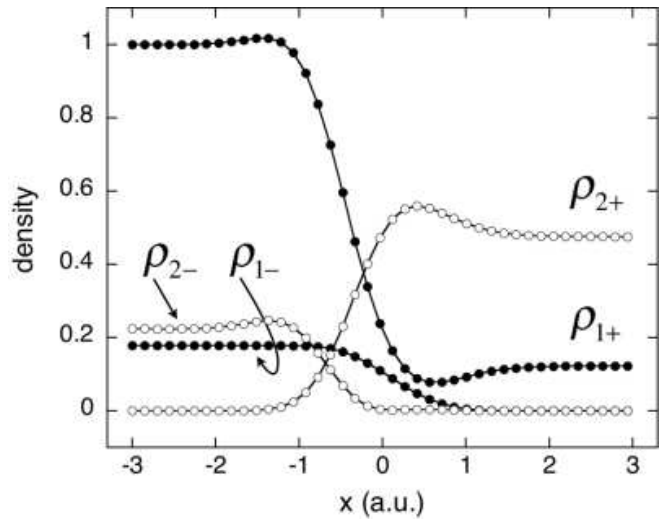}
        \caption{Component wave densities as a function of position, for the
                 two-surface pure coupling system (Sec.~\ref{pcsys}). Circles indicate
                 trajectory grid points at the final time, for both diabatic states,
                 1 (filled circles) and 2 (open circles).}
        \label{pcall}
\end{figure}

\begin{figure}
\includegraphics[scale=0.85]{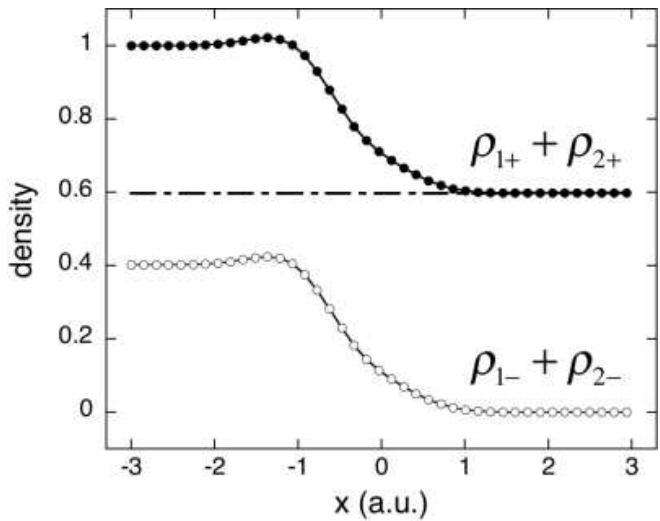}
        \caption{Total transmitted (filled circles) and reflected (open circles) component
                 wave densities (summed over all diabatic states) as a function of position,
                 for the two-surface pure coupling system (Sec.~\ref{pcsys}). Circles indicate
                 trajectory grid points at the final time, $t = t_{\text{max}}$. Dot-dashed line
                 indicates difference between the two curves---found to be a constant, equal to
                 the total transmission probability.}
        \label{pcsum}
\end{figure}

The four resultant converged component densities, $\rho_{1\pm}(x)$  and $\rho_{2\pm}(x)$,
are plotted in Fig.~\ref{pcall} as a function of x. All four curves are smooth and
interference-free, as desired, and satisfy the appropriate boundary conditions of
\eq{bcs}. From \eq{Prefltrans}, the various reflection and transmission probabilities
are found to be as follows: $P_1^{\text{refl}}= 0.17886$; $P_2^{\text{refl}}= 0.22382$;
$P_1^{\text{trans}}= 0.12194$; $P_2^{\text{trans}}= 0.47537$. In Fig.~\ref{pcsum}, the two
transmitted wave component densities, $\rho_{1+}(x)$ and $\rho_{2+}(x)$, are summed
together to form $\rho_+(x)$, and compared with the corresponding sum for the
reflected components. As per the discussion at the end of Sec.~\ref{multisurftheory},
the two summed curves are indeed found to be identical---apart from a constant
shift, represented by the dot-dashed line.

\subsubsection{Tully Model 1}
\label{model1}

The Tully models were introduced in the early 1990's by John Tully
and coworkers,\cite{tully90} to serve as benchmark numerical
applications for investigating various processes involving
electronic transitions.

The first Tully model is a simple avoided crossing system that consists of two diabatic
ramp potentials---one an uphill ramp, and the other a downhill ramp---which cross each
other symmetrically. For consistency and convenience, we have replaced Tully's original
piece-wise-exponential ramps with an analogous tanh functional form. The modified
Tully Model 1 potentials are therefore as follows:
\ea{
     V_{11}(x) & = & \of{{V_{R}-V_{L} \over 2}} \tanh \sof{\beta (x-x_0)} \nonumber \\
     V_{22}(x) & = & \of{{V_{L}-V_{R} \over 2}} \tanh \sof{\beta (x-x_0)} \nonumber \\
     V_{12}(x) & = & V^{12}_0 \exp[-\alpha (x-x_0)^2], \label{model1pot}
}
with parameters $V_R = -V_L = 0.01$ hartree, $x_0=0$, $\beta =
1.2$ a.u., $V^{12}_0 = 0.005$ hartree, and $\alpha = 1$ a.u.

\begin{figure}
\includegraphics[scale=0.85]{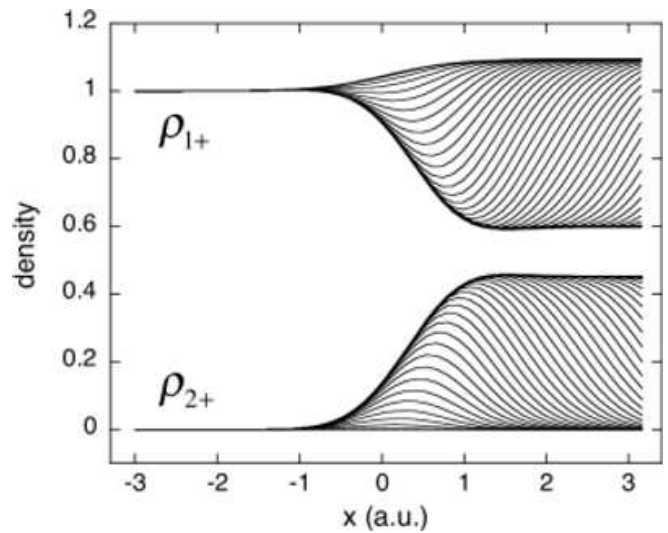}
        \caption{Component transmitted wave densities as a function of position, and for a variety
                 of times, $t$, for the two-surface Tully Model 1 system (Sec.~\ref{model1}). The
                 upper family of curves represent $\rho_{1+}(x)$ at different times, whereas the
                 lower family of curves represent $\rho_{2+}(x)$. The highest $\rho_{1+}$
                 curve is the $t=0$ initial value (i.e. the WKB approximation), whereas the lowest
                 $\rho_{2+}(x)=0$ curve represents zero initial probability on diabatic state 2.
                 Over time, probability is transferred from diabatic state 1 to 2 as indicated,
                 via intersurface coupling.}
        \label{model1trans}
\end{figure}

The diagonal potentials, $V_{ii}(x)$, are no longer symmetric,
necessitating the use of \eq{Pdotgeneral} rather than
\eq{Pdotspecial}. Due to the monotonicity of these potential curves,
we take $\Veff_i(x)=V_{ii}(x)$, which results in standard classical
trajectories for the bipolar CPWM dynamics. Note that there are now
potential couplings that can induce direct transitions between any
pair of $\Ppmi 1$ and $\Ppmi 2$ components. However, the energy
value chosen, $E = 0.11$ hartree ($\approx 24 142\,\text{cm}^{-1})$,
is so much larger than the potential energy range as to ensure that
there is negligible reflection. On the other hand, a substantial
amount of intersurface coupling is obtained (Fig.~\ref{model1trans}).

Numerical calculations were performed as in Sec.~\ref{pcsys}, using
the following parameter values: $N=50$; $x_{L/R} = \mp 3$ a.u.;
$t_{\text{max}} = 1000$ a.u.; $\epsilon = 10^{-6}$. An ``animation
plot'' of the resultant time-dependent transmitted wave densities,
$\rho_{1+}(x,t)$ and $\rho_{2+}(x,t)$, is presented in
Fig.~\ref{model1trans}. Each ``contour line'' is actually snapshot of
a given component density at a particular time, $t$. For
$\rho_{1+}$, the uppermost curve is $\rho_{1+}(x)$ at the initial
$t=0$, whereas the lowermost $\rho_{1+}$ curve represents the $t\ra
\infty$ limit. The opposite relation holds for the $\rho_{2+}$
curves. From the figure, it is clear that the time evolution is
smooth and well-behaved, and converges to the large-time limit
quickly. The final curves represent essentially 100\% transmission
(the reflected wave densities are negligibly small, as expected),
split roughly equally between the two curves as
$P_1^{\text{trans}}= 0.55016$ and $P_2^{\text{trans}}= 0.44983$.

\begin{figure}
\includegraphics[scale=0.85]{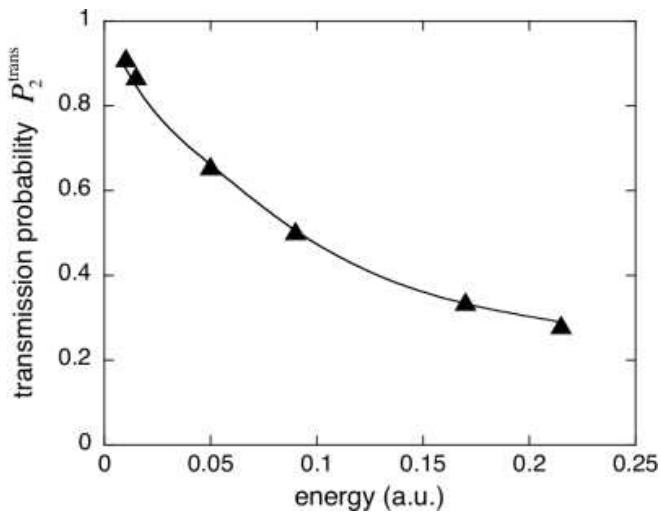}
        \caption{Computed transmission probabilities $P_2^{\text{trans}}$ vs energy $E$,
                 for the two-surface Tully Model 1 system (Sec.~\ref{model1}). Filled
                 triangles indicate present results, as obtained using the classical
                 trajectory CPWM bipolar decomposition; solid curve indicates quantum
                 results of Herman.\cite{herman05}}
        \label{model1energy}
\end{figure}

In addition to the $E = 0.11$ hartree case above, bipolar CPWM
calculations were repeated and reconverged for a wide range of other
energy values, to an accuracy of $10^{-4}$ or better.
Computed transmission
probabilities for $P_2^{\text{trans}}$ are presented in
Fig.~\ref{model1energy}, and found to be in excellent agreement
with the quantum results of Herman.\cite{herman05}

\subsubsection{Tully Model 2}
\label{model2}

The second Tully model is a more complex, double avoided crossing
system. The two diabatic curves consist of a symmetric well, and a
constant energy curve which cuts across the well, giving rise to
avoided crossings at the two intersections.  These in turn give rise
to quantum interference effects.\cite{tully90,wan00} The potentials are
\ea{
     V_{11}(x) & = & 0 \nonumber \\
     V_{22}(x) & = & - V_0 \exp[-\beta (x-x_0)^2] + E_0 \nonumber \\
     V_{12}(x) & = & V^{12}_0 \exp[-\alpha (x-x_0)^2], \label{model2pot}
}
with parameters $V_0 = 0.10$ hartree, $x_0=0$, $\beta = 0.28$
a.u., $E_0 = 0.05$ hartree, $V^{12}_0 = 0.015$ hartree, and $\alpha
= 0.06$ a.u.

The diagonal potentials are symmetric, and therefore amenable to
uniform trajectories. However, the energies involved are so large and
``classical-like'' that it is better to use standard classical
trajectories, $\Veff_i(x)=V_{ii}(x)$. Note that $V_{22}$ describes
a potential well, thus ensuring that there will be no turning points.
The large energies also ensure that reflection is negligible,
but again, there is substantial intersurface coupling. Depending on the energy,
the following range of parameters were used to achieve a convergence of
$10^{-2}$ or better: $N=200$--$300$; $x_{L/R} = \mp 8$ a.u.;
$t_{\text{max}} = 1000$--$2000$ a.u.; $\epsilon = 10^{-4}$.

\begin{figure}
\includegraphics[scale=0.85]{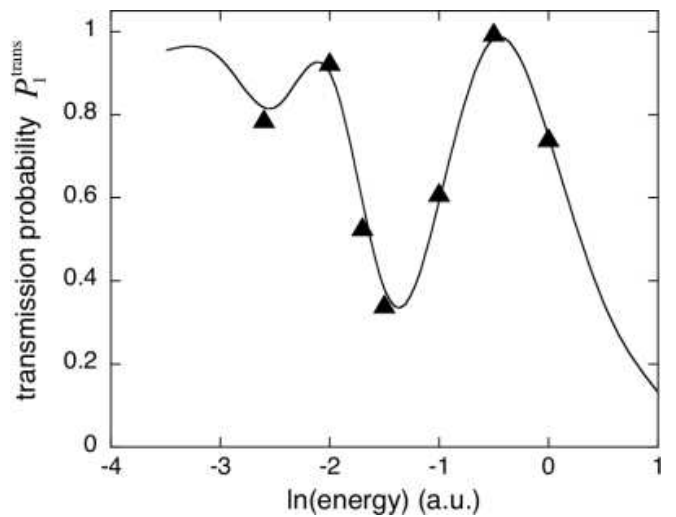}
        \caption{Computed transmission probabilities $P_1^{\text{trans}}$ vs logarithm of
                 energy $E$, for the two-surface Tully Model 2 system (Sec.~\ref{model2}).
                 Filled triangles indicate present results, as obtained using the classical
                 trajectory CPWM bipolar decomposition; solid curve indicates quantum
                 results of Herman.\cite{herman05} Note the St\"uckelberg oscillations.}
        \label{model2energy}
\end{figure}

Quantum interference associated with the two different surface
pathways manifests as two distinct types of ``St\"uckelberg
oscillations,''\cite{tully90,wan00,nikitin,drake} both associated
with the St\"uckelberg phase
difference,\cite{drake,poirier07bohmIVfootnote}
\eb
     \Dlt \Phi_{12} \approx \of{{1\over \hbar}} \int \sof{\sqrt{2m(E-V_{11})}-
           \sqrt{2m(E-V_{22})}} dx'. \label{stuck}
\ee
Evaluating \eq{stuck} as a definite integral over the relevant range
of $x'$ yields
oscillations in $P_1^{\text{trans}}$ as a function of $E$. The
computed transmission probabilities over a range of $E$ values
are presented in Fig.~\ref{model2energy}; these are indeed
oscillatory, and also in  good agreement with \Ref{herman05}.

\begin{figure}
\includegraphics[scale=0.85]{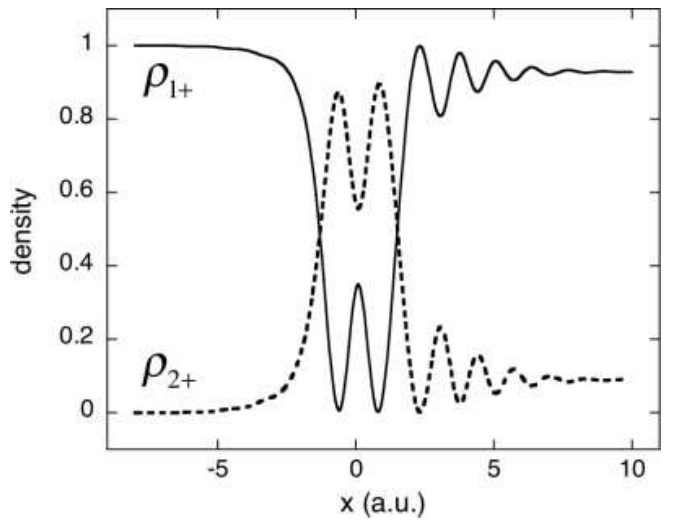}
        \caption{Component transmitted wave densities, $\rho_{1+}$ (solid) and $\rho_{2+}$
                 (dashed), as a function of position, for the two-surface Tully Model 2
                 system (Sec.~\ref{model2}). The oscillatory transfer of probability is
                 a type of St\"uckelberg phenomenon.}
        \label{model2trans}
\end{figure}

The second type of St\"uckelberg oscillation is obtained by
integrating \eq{stuck} to the indefinite limit, $x$, which can
give rise to oscillations in $\rho_i(x) \approx \rho_{i+}(x)$.
Oscillations are indeed observed in the converged transmitted wave densities
for the $E=\exp(-2) \approx 0.13533$ hartree case considered, as presented in
Fig.~\ref{model2trans}. The asymptotic oscillation wavelength as
predicted by \eq{stuck} is found to be $1.31$ a.u., in good agreement
with the figure.

\section{CONCLUSION}
\label{conclusion}

From a numerical perspective, the straight fourth-order Runge-Kutta
bipolar CPWM algorithm, incorporating the numerical refinements
of \Ref{poirier07bohmalg}, is found in Sec.~\ref{singleapp} to satisfy
the demands of numerical stability, robustness, efficiency,
and minimal user intervention. In particular, this algorithm is remarkably
efficient at both high and low levels of absolute accuracy, and
also for very high precision calculations such as Table~\ref{pottab}
Column III. We believe this algorithm to be the fastest available for
computing stationary scattering states of single-surface 1D systems with
predetermined boundary conditions, and advocate its use as a ``black-box''
method that can be applied to virtually any such application.
That the algorithm also generalizes to {\em multisurface} 1D
applications in straightforward fashion (Sec.~\ref{multisurfalg}),
is also significant. We note that for systems such as Tully Model 2,
exhibiting substantial intersurface but little $(+\!\leftrightarrow\!-)$
interference effects, the bipolar CPWM in its present incarnation necessarily
undergoes a noticeable drop in efficiency, as the St\"uckelberg oscillations
must manifest in the component field functions. However, Fig.~\ref{model2trans}
suggests that it may be possible to generalize the theory for arbitrary electronic
representations (i.e. neither diabatic nor adiabatic) in such a way as to avoid
interference oscillations altogether in such cases. In any event, well-documented
stand-alone fortran codes for the one- and two-surface algorithms considered here
are available from the authors on request.

Note that although little explicit connection with conventional ``Bohmian mechanics''
{\em per se} has been made in this work---i.e. of ``polar'' (amplitude-phase)
decompositions of the wavefunction components, and quantum potentials---implicitly,
it is precisely this connection that enables the algorithms used here to work so well,
as discussed in detail in the previous publications, notably Refs.~3 and~4. Of
course, it is possible to recast \eqs{Pdotgeneral}{Pdotspecial} in polar form,
yielding more explicitly ``Bohm-like'' results. Indeed, this is the usual
convention in the prior semiclassical literature,\cite{froman} and we have also
done this. However, this results in more complicated evolution equations that
obscure aspects of the physics which we wish to emphasize in this work.
Note that from a {\em numerical} perspective however, we use {\em both}
polar and non-polar representations at different stages in the algorithm,
transforming between them as described in \Ref{poirier07bohmalg} [Eq.~(9)].

From a theoretical perspective, the multisurface generalization of the
F approach with generalized classical trajectories
(Sec.~\ref{multisurftheory}) presents some very intriguing aspects,
with possible significance beyond the scope of the present paper.
To begin with, the flux relations of Eqs.~(\ref{fluxrel})
through~(\ref{contcond}) provide a natural, but non-trivial,
multi-surface-like interpretation to the flow of probability among all
$2f$ of the ``components,'' $\Ppmi i$. Even more compelling, however,
is the form of the dynamical equations of \eq{Pdotspecial}, which above all
serves to highlight the essential sameness of the two types of
transitions---intersurface and $(+\!\leftrightarrow\!-)$---now treated
together within a single unified framework. That off-diagonal diabatic
potentials induce transitions from one surface to another has always
been known; now we find that in similar fashion, {\em diagonal} potentials
also induce transitions, between transmitted and reflected components of the
wavefunction associated with a given diabatic state.

The above picture can lead one in a variety of interesting directions.
Traditional TSH methods, for instance,\cite{tully71,tully90}
utilize off-diagonal coupling\cite{poirier07bohmIVfootnote}
as a means of inducing trajectory hops from one surface to another.
The present work suggests that a similar TSH scheme could be applied
to effect transitions from reflected to transmitted wavepackets
(even for single-surface calculations), thereby naturally introducing
interference and other effects that might otherwise be problematic
in a traditional TSH context. Such a method would first require
that a non-stationary, {\em wavepacket} generalization of the
present work be developed; indeed, this will serve as the focus
of the next publication in the series, with the subsequent paper
addressing multidimensional applications. Note that both of these
future papers will more directly emphasize the link with Bohmian
mechanics---restoring the quantum potential, for example, which is
required for wavepackets, but not essential for stationary state applications.
As a bit of foreshadowing to motivate the present work, we comment that the
multidimensional generalization still requires only {\em two} components
per electronic state (regardless of system dimensionality), and can
accommodate standard Jacobi-type coordinate representations with arbitrary
curvilinear reaction paths.

Finally, we briefly mention two other possible areas for future development.
First, the present theory is restricted to potentials that are asymptotically
flat in {\em both} asympotes, $x\ra\infty$ and $x\ra -\infty$. It would be
useful to generalize for potentials that are singular, or otherwise diverge
in one asymptote or the other. Second, the intriguing result from
Table~\ref{pottab} Row 6 that $(\tshift/\Dlt)$ is always on the order of one
suggests that simply {\em setting} this value to one would not adversely
affect numerical efficiency too severely. Indeed, such a modification would
lead to some decided numerical advantages, such as the use of fixed grids
which would avoid the need for intrasurface (but not intersurface)
interpolation. The resultant grid densities would be larger, however,
and for this reason, such an approach might be substantially less
efficient for high dimensionalities.

\begin{acknowledgments}

This work was supported by a grant from The Welch Foundation (D-1523).
Jason McAfee is also acknowledged for his aid in converting this manuscript to an electronic format suitable for the arXiv preprint server.

\end{acknowledgments}

%
%

%
%
%



\begin{thebibliography}{10}
\bibitem{poirier07bohmIVfootnote}
Strictly speaking adiabatic potentials should be used, but this distinction is
  immaterial for the present discussion.
\end{thebibliography}
\end{document}